\documentclass[prd,superscriptaddress,nofootinbib,amsfonts,notitlepage]{revtex4-2}
\usepackage[utf8]{inputenc}
\usepackage[T1]{fontenc}
\usepackage{hyperref}
\usepackage{graphicx,bm,amsmath,color}
\usepackage{bbold}
\usepackage{wasysym}
\usepackage{verbatim}
\usepackage{amssymb}
\usepackage{epstopdf}
\usepackage{animate}
\usepackage{xmpmulti} 
\usepackage{centernot}
\usepackage{multirow}
\usepackage{tikz}
\usepackage{graphicx}
\usepackage{float}
\usepackage{mathtools}
\usepackage{comment}
\usetikzlibrary{arrows}
\usetikzlibrary{decorations.pathreplacing}

\makeatletter

\newcommand{\be}{\begin{equation}}
\newcommand{\ee}{\end{equation}}
\newcommand{\beq}{\begin{eqnarray}}
\newcommand{\eeq}{\end{eqnarray}}
\newcommand{\ba}{\begin{align}}
\newcommand{\ea}{\end{align}}
\newcommand{\red}[1]{\textcolor{red}{#1}}

\begin{document}

\title{Space-time thermodynamics in momentum dependent geometries}

\author{Goffredo Chirco}
\email{goffredo.chirco@unina.it}
\affiliation{Dipartimento di Fisica ``Ettore Pancini'', Università di Napoli Federico II, Napoli 80125, Italy;}
\affiliation{INFN, Sezione di Napoli, Napoli 80125, Italy}

\author{Stefano Liberati}
\email{liberati@sissa.it}
\affiliation{SISSA, Via Bonomea 265, 34136 Trieste, Italy and INFN, Sezione di Trieste;}
\affiliation{ IFPU - Institute for Fundamental Physics of the Universe, Via Beirut 2, 34014 Trieste, Italy}

\author{José Javier Relancio}
\affiliation{Departamento de Física, Universidad de Burgos, 09001 Burgos, Spain;}
\affiliation{Departamento de F\'{\i}sica Te\'orica and Centro de Astropartículas y F\'{\i}sica de Altas Energ\'{\i}as (CAPA),
Universidad de Zaragoza, Zaragoza 50009, Spain
}
\email{jjrelancio@ubu.es}

\begin{abstract}
%In a quantum gravity theory, at very high energies and very small scales, we expect the structure of spacetime to be radically modified with respect to the classical notion of special and general relativity.
A possible way to capture the effects of quantum gravity in spacetime at a \emph{mesoscopic} scale, for relatively low energies, is through an energy dependent metric, such that particles with different energies probe different spacetimes. In this context, a clear connection between a geometrical approach and modifications of the special relativistic kinematics has been shown in the last few years. In this work, we focus on the geometrical interpretation of the relativistic deformed  kinematics present in the framework of doubly special relativity, where a relativity principle is present. In this setting, we study the effects of a momentum dependence of the metric for a uniformly accelerated observer. We show how the local Rindler wedge description gets affected by 
%the $k^2$ dependence of 
the proposed observer dependent metric, while the local Rindler causal structure is not, leading to a standard local causal horizon thermodynamic description. For the proposed modified metric, we can reproduce the derivation of Einstein's equations as the equations of state for the thermal Rindler wedge. The conservation of the Einstein tensor leads to the same privileged momentum basis obtained in other works of some of the present authors, so supporting its relevance. 
\end{abstract}

\maketitle

\section{Introduction}
One of the main open problems of theoretical physics is the unification of general relativity (GR) and quantum field theory (QFT), or equivalently, the formulation of a quantum gravity theory (QGT). %One of the possible issues impeding us to such construction resides in the role that spacetime plays in them: it is a dynamical variable in GR, and a static frame in QFT.
The search for a QGT has lead to several, very diverse, theoretical frameworks, such as string theory~\cite{Mukhi:2011zz,Aharony:1999ks,Dienes:1996du}, loop quantum gravity~\cite{Sahlmann:2010zf,Dupuis:2012yw}, causal dynamical triangulations \cite{Loll_2019}, causal set theory~\cite{Wallden:2013kka,Wallden:2010sh,Henson:2006kf}, to name a few.  In (almost) all of them, a minimum length scale arises~\cite{Gross:1987ar,Amati:1988tn,Garay1995},  which is heuristically associated with the Planck length $\ell_P \sim 1.6\times 10^{-33}$\,cm (or Planck mass $M_P \sim 1.22\times 10^{19}$\,GeV). Such a small length (high-energy scale) is expected to separate the regime where spacetime displays a classical geometric structure from the one where it develops its quantum nature.

To date, none of the aforementioned theories have proved fully satisfactory, in the sense that they do not provide a well established phenomenology. Nonetheless, much work in quantum gravity phenomenology has been produced recently starting from the idea that the effects of quantum spacetime can be captured, beside a fundamental QGT, by considering the low-energy limit of a given scenario for the UV completion of the standard model and general relativity, and seeking for observable implications~\cite{Addazi:2021xuf}. 

One of these commonly entailed scenarios is the one in which the usual local  Poincaré symmetry of spacetime happens to be emergent at long wavelengths/low energies~\cite{Amelino-Camelia:2000cpa,Amelino-Camelia:2000stu}. In particular, this is achieved by modifying the special relativistic kinematics with the introduction of a high-energy scale often, but not necessarily, identified with the Planck scale. There are two different possibilities that can be explored in this sense. One can consider that local Lorentz symmetry is violated for energies comparable to this scale, leading to  the framework of Lorentz invariance violation (LIV)~\cite{Colladay:1998fq,Kostelecky:2008ts,Mattingly:2005re,Liberati:2013xla}. In this case, there is a breakdown of the relativity principle that characterizes special relativity (SR), and therefore, a privileged observer appears. A covariant description of this phenomenology requires the introduction of an aether field setting such preferred system of reference so leading to modified gravity theories such as Einstein--Aether theory~\cite{Jacobson:2000xp} and Ho\v{r}ava gravity~\cite{Horava:2009uw}.

A different scenario consists in exploring the possibility that the very Lorentz symmetry is deformed. This is studied in doubly (or deformed) special relativity (DSR) theories~\cite{Amelino-Camelia:2000cpa}. In DSR,  the Einsteinian relativity principle is generalized by adding to the light speed scale an extra (high-)energy scale. 

Due to the different implementation of space-time symmetries within the two theories, the modification of the special relativistic  kinematics are also different. While in LIV effective field theories the only modification arises in a modified dispersion relation, in DSR, in addition to this possible ingredient (this is not the case for the well known kinematics of the classical basis of $\kappa$-Poincaré~\cite{Borowiec2010}), there is a deformed conservation law for momenta. This last ingredient is indispensable in order to have a relativity principle. Indeed, this is ensured by the existence of some Lorentz transformations, in the one- and two-particle systems, which make the two previous ingredients compatible.

During the last years, a clear connection between a curved momentum space, which naturally leads to a momentum dependent geometry, and these relativistic deformed kinematics of $\kappa$-Poincaré, has been established~\cite{Kowalski-Glikman:2002oyi,Kowalski-Glikman:2003qjp,Freidel:2011mt,Banburski:2013jfa}.  For example, it was suggested in~\cite{Magueijo:2002xx} that DSR could be the outcome of an energy (rainbow) spacetime, thereby showing a clear connection between a momentum dependent spacetime and quantum gravity.  

More recently,  it was rigorously shown that all the ingredients of a relativistic deformed kinematics characterizing DSR can be obtained from a maximally symmetric momentum space~\cite{Carmona:2019fwf}. In particular,   $\kappa$-Poincaré kinematics  \cite{Lukierski:1991pn,Lukierski:1993df,Lukierski:1992dt,Lukierski:2002df}  can be obtained from a de Sitter momentum space\footnote{By a similar procedure one can generalize this construction in order to obtain the kinematics of Snyder~\cite{Battisti:2010sr} and hybrids models~\cite{Meljanac:2009ej}, and to the case of anti de Sitter.}: the isometries (translations and Lorentz isometries) and the squared distance of the metric, are identified with the deformed composition law, deformed Lorentz transformations, and deformed dispersion relation, respectively (the last two facts were previously contemplated in Refs.~\cite{AmelinoCamelia:2011bm,Lobo:2016blj}). In~\cite{Relancio:2020zok}, the proposal of~\cite{Carmona:2019fwf} was generalized so as to allow the metric to describe a curved spacetime, leading to a metric in the cotangent bundle depending on all the phase-space variables. This is a generalization of previous works in the literature, the so-called generalized Hamilton spaces, in which a metric that depends on the velocities (Finsler geometries)~\cite{Girelli:2006fw,Amelino-Camelia:2014rga,Lobo:2016xzq} and momenta (Hamilton geometries)~\cite{Barcaroli:2015xda,Barcaroli:2016yrl,Barcaroli:2017gvg} were considered\footnote{It is worth mentioning that a Finsler/momentum dependent spacetime can be introduced also as an alternative description of Lorentz breaking physics~\cite{Barcelo:2001cp,Kostelecky:2011qz,Stavrinos:2016xyg,Hasse:2019zqi}.}. 

Also in DSR scenarios, the study of the propagation and interaction of particles considering a curvature in both
momentum and space-time spaces was carried out in
Ref.~\cite{Cianfrani:2014fia}. In that paper an action with some nonlocal variables (defined by the space-time tetrad) is considered,
allowing one to generalize the relative locality action~\cite{AmelinoCamelia:2011bm}
when a curvature in spacetime is present.

In the DSR community there is an open debate about the possibility that different kinematics of $\kappa$-Poincaré, associated to different bases in Hopf algebras~\cite{KowalskiGlikman:2002we}, or equivalently, different choices of momentum coordinates in a de Sitter momentum space, could represent different physics~\cite{AmelinoCamelia:2010pd}.  This is due to the existence of an ambiguity about the definition of momentum variables associated to physical measurements. While in the flat spacetime/curved momentum space geometrical setup there is such an ambiguity, this degeneracy is broken when considering that both spacetime and momentum space  are curved~\cite{Relancio:2020rys}. Indeed, in~\cite{Pfeifer:2021tas} it was shown that only Lorentz covariant metrics are allowed in the proposed geometrical scheme when lifting the deformed symmetries to a curved spacetime, and in~\cite{Relancio:2020rys} it was proposed a way to select a ``physical'' basis by imposing the conservation of the Einstein tensor. Remarkably, the so selected special basis was shown to be also the one in which all the definitions of surface gravity are equivalent for Killing horizons~\cite{Relancio:2021asx}, as they are in standard GR~\cite{Cropp:2013zxi}.  

In this context, an interesting question is, if starting from a cotangent bundle geometry, it is possible to get the Einstein equations by adopting a thermodynamical derivation as that proposed by Jacobson in 1995~\cite{Jacobson:1995ab} and based on the application of the Clausius equation on a local Rindler wedge. A Rindler spacetime in the DSR context was considered not long ago in~\cite{Arzano:2017opw}, with regards to a twist deformation of the algebra describing a relativistic deformed kinematics. That paper considered the modes of massless scalar fields using a deformed (momentum dependent) measure. Unfortunately, a QFT based on DSR is still missing, albeit this issue has been the subject of several investigations over the past two decades~\cite{Kosinski:2001ii,Govindarajan:2009wt,Poulain:2018two,Arzano:2020jro,Lizzi:2021rlb,Franchino-Vinas:2022fkh}. Such a dynamical theory, going beyond the usual relativistic QFT, would have to take into account a relativistic deformed kinematics, so that, while the usual QFT has at its base the Poincaré symmetry, a DSR QFT would have to rest on a deformed symmetry group characterizing a relativistic deformed kinematics. 

Keeping this in mind, in this paper we will apply the geometrical interpretation of DSR developed in \cite{Relancio:2020zok,Relancio:2020mpa,Relancio:2020rys,Relancio:2021asx,Pfeifer:2021tas} to an accelerated observer. We will see that the usual derivation followed in GR for constructing a Rindler spacetime is compatible with our setting, allowing us to consider at the same time a curved spacetime and a momentum dependent metric compatible with a relativistic deformed kinematics. More precisely, we shall here derive the Einstein equations from a thermodynamical point of view (instead of considering the proposal of~\cite{miron2001geometry} used in~\cite{Relancio:2020rys}). Starting from the restricted metrics obtained in~\cite{Pfeifer:2021tas}, we will find that the conservation of the energy-momentum tensor leads to the same Einstein equations of GR, replacing the metric, the Ricci tensor, and the scalar of curvature, by its generalizations in the cotangent bundle, which agrees with the proposal of~\cite{miron2001geometry}. The only difference with respect the usual equations is 
a term proportional to the mass of the particle probing spacetime, which, as we will see, can be regarded as a cosmological constant. With these equations,  the   energy-momentum tensor is only conserved in the privileged momentum basis previously found in~\cite{Relancio:2020rys,Relancio:2021asx}.

The structure of the paper is as follows. In Sec.~\ref{sec:geometrical_intro} we make a brief review of the main ingredients of a cotangent bundle geometry and about our  construction of a metric in phase space accounting for a relativistic deformed kinematics in curved spacetimes. The derivation of the Rindler (momentum dependent) spacetime is carried out in Sec.~\ref{sec:rindler}. We check that our geometrical construction is compatible with our generic method of building a cotangent bundle metric. In Sec.~\ref{sec:einstein} we discuss the Einstein equations in our scheme by regarding a thermodynamical point of view. Finally, we end with the conclusions in Sec.~\ref{sec:conclusions}.

\section{Cotangent bundle in a nutshell}
\label{sec:geometrical_intro}
In this section, we review the main geometrical ingredients in the metric description of the phase-space cotangent bundle as first introduced in~\cite{miron2001geometry}, as well as the main results from the previous literature concerning the set up of a relativistic deformed  kinematics in a curved space-time background.  

\subsection{Main properties of the geometry in the cotangent bundle}

As mentioned in the introduction, there are several works in the literature trying to connect a momentum dependent geometry with a relativistic deformed kinematics. The most natural way to consider a curved metric in momentum space is by regarding the cotangent lift $T^*M$ of a space-time manifold $M$. For a given base manifold $M$, with coordinates $x$, one can construct its cotangent lift $T^*M$, with coordinates $(x,k)$. Therefore, the cotangent manifold has 8 dimensions when considering that the base manifold (the spacetime where particles propagates) has 4 dimensions, and generically depends on the space-time point. For a fixed $x$, the set of all $k$'s is called the fiber, and represents locally the momentum space. It is important to remark that these variables, $x$ and $k$, are canonically conjugated under the structure of Poisson brackets,
\begin{equation}
    \lbrace x^\mu,k_\nu \rbrace\,=\, \delta^\mu_\nu\,. 
\end{equation}

\begin{figure}
\centering
\includegraphics[width=14cm]{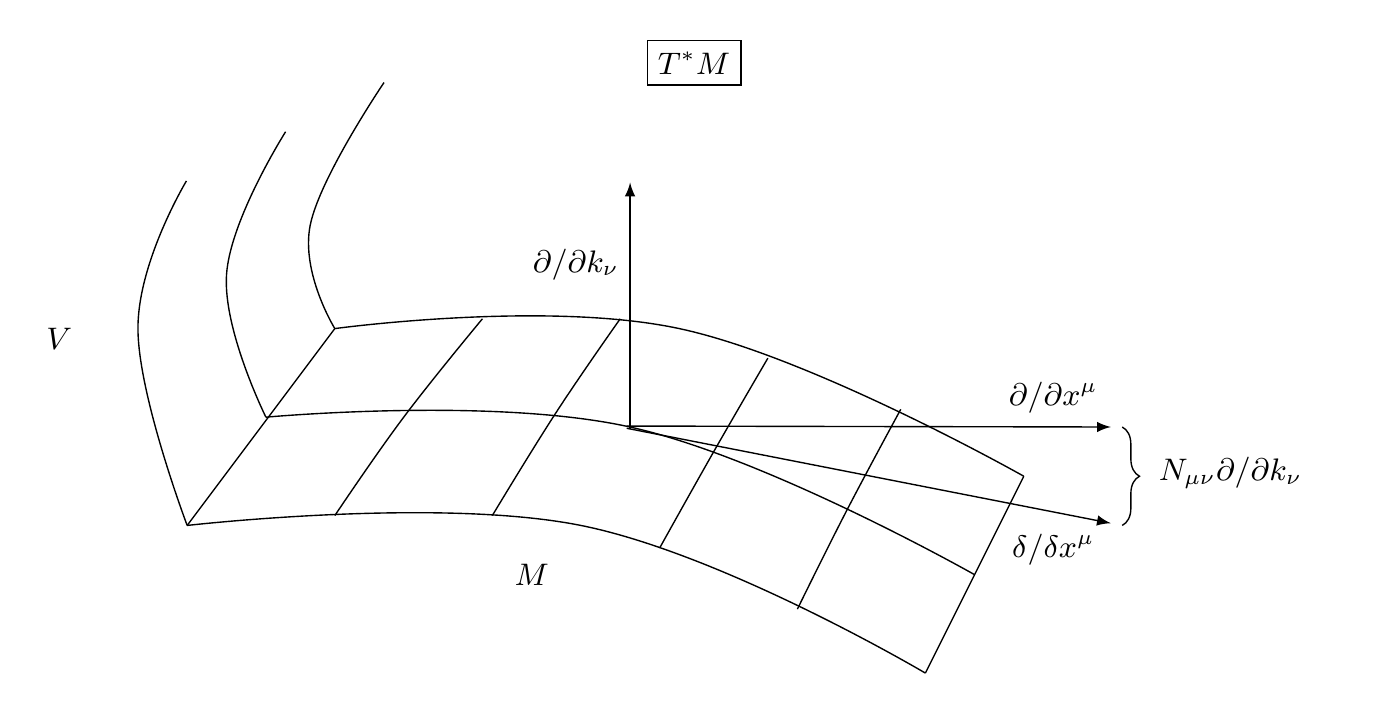}
\caption{A visualization of the cotangent bundle structure~\cite{Relancio:2020rys}} 
\label{figure_cotangent}
\end{figure}

On the cotangent bundle manifold, associating to each point $u \in T^*M$ (i.e. a point in phase space $(x,k)$) the fiber $V_u$ (all the points with fixed $x$ but different $k$), one can obtain the so-called vertical distribution, $V: u \in T^*M\rightarrow V_u \subset T_u T^*M$ with dimension $n$, which is generated by $\partial/\partial k$. Here, $T_u T^*M$ is the tangent space of the manifold $T^*M$. As it is shown in Fig~\ref{figure_cotangent}, given a point on the cotangent bundle, one can construct the vertical distribution (the fiber). Note that in the figure the fiber is unidimensional for the sake of simplicity, but in fact it has the same dimensions as the base manifold. 

In this setting, one can define a  line element in the whole cotangent bundle~\cite{miron2001geometry}
\begin{equation}
\mathcal{G}\,=\, g_{\mu\nu}(x,k) dx^\mu dx^\nu+g^{\mu\nu}(x,k) \delta k_\mu \delta k_\nu\,,
\label{eq:line_element_ps} 
\end{equation}
where 
\begin{equation}
\delta k_\mu \,=\, d k_\mu - N_{\nu\mu}(x,k)\,dx^\nu\,, 
\end{equation}
being $N_{\nu\mu}(x,k)$ the so-called nonlinear connection coefficients,  and $g_{\mu\nu}(x,k)$ is the cotangent-bundle metric tensor which we shall discuss in detail later on. We can see that there are two different types of privileged curves: ones for which $x$ is fixed, leading to movements along the momentum space (fiber), and another ones for which $\delta k=0$. The latter
condition in Eq.~\eqref{eq:line_element_ps}  leads to a GR-like line element on which it is possible to define geodesic motion via simple generalization of the geodesic equation, that takes the form~\cite{miron2001geometry}
\begin{equation}
\frac{d^2x^\mu}{d\tau^2}+{H^\mu}_{\nu\sigma}(x,k)\frac{dx^\nu}{d\tau}\frac{dx^\sigma}{d\tau}\,=\,0\,,
\label{eq:horizontal_geodesics_curve_definition}
\end{equation} 
where 
\begin{equation}
{H^\rho}_{\mu\nu}(x,k)\,=\,\frac{1}{2}g^{\rho\sigma}(x,k)\left(\frac{\delta g_{\sigma\nu}(x,k)}{\delta x^\mu} +\frac{\delta g_{\sigma\mu}(x,k)}{\delta  x^\nu} -\frac{\delta g_{\mu\nu}(x,k)}{\delta x^\sigma} \right)\,,
\label{eq:affine_connection_st}
\end{equation} 
is the (metrical) affine connection of spacetime, and \begin{equation}
\frac{\delta}{\delta x^\mu}\, \doteq \,\frac{\partial}{\partial x^\mu}+N_{\nu\mu}(x,k)\frac{ \partial}{\partial k_\nu}\,.
\label{eq:delta_derivative}
\end{equation}
Here, $\tau$ plays the role of the proper time or of the affine parametrization, depending on whether one is considering a massive or a massless particle, respectively.

The choice of the nonlinear connection coefficients   $N_{\nu\mu}(x,k)$ is not unique, but, as shown in~\cite{miron2001geometry}, there is one and only one choice of nonlinear connection coefficients that leads to a space-time affine connection which is metric compatible and torsion free.
When the metric does not depend on the momentum, as it is standard in GR, the coefficients of the nonlinear connection are given by
\begin{equation}
N_{\mu\nu}(x,k)\, = \, k_\rho\, {\Gamma^\rho}_{\mu\nu}(x)\,,
\label{eq:nonlinear_connection}
\end{equation} 
where $ \Gamma^\rho_{\mu\nu}(x)$   are the usual  Christoffel symbols of GR. On the other hand, when the metric does not depend on the space-time coordinates, $N_{\mu\nu}(x,k)$ generically vanishes.

The space-time covariant derivative of a tensor can be as well be defined as~\cite{miron2001geometry}
\begin{equation}
\begin{split}
T^{\alpha_1 \ldots\alpha_r}_{\beta_1\ldots\beta_s;\mu}(x,k)\,&=\,\frac{\delta T^{\alpha_1 \ldots\alpha_r}_{\beta_1\ldots\beta_s}(x,k)}{\delta x^\mu}+T^{\lambda \alpha_2 \ldots\alpha_r}_{\beta_1\ldots\beta_s}(x,k){H^{\alpha_1}}_{\lambda \mu}(x,k)+\cdots+T^{\alpha_1 \ldots \lambda}_{\beta_1\ldots\beta_s}(x,k){H^{\alpha_r}}_{\lambda \mu}(x,k)\\
&-T^{\alpha_1 \ldots \alpha_r}_{\lambda \beta_2\ldots\beta_s}(x,k){H^{\lambda}}_{\beta_1 \mu}(x,k)-\cdots-T^{\alpha_1 \ldots \alpha_r}_{\beta_1\ldots \lambda}(x,k){H^{\lambda}}_{\beta_s \mu}(x,k)\,.
\label{eq:cov_dev_st}
\end{split}
\end{equation} 
Also, it is shown that, given a metric, there is always a symmetric nonlinear connection coefficients (aka $N_{\mu\nu}$) leading to the affine connections in spacetime such that the covariant derivative of the metric vanishes:
\begin{equation}
g_{\mu\nu;\rho}(x,k)\,=\,0\,.
\label{eq:covariant_derivative_2}
\end{equation}

In order to study the properties of the horizon, we need to know how the Lie derivative is deformed in this context. In~\cite{Barcaroli:2015xda,Relancio:2020zok} the modified Killing equation for a metric in the cotangent bundle was derived
\begin{equation}
\frac{\partial g_ {\mu\nu}(x,k)}{\partial x^\alpha} \chi^\alpha  -\frac{\partial g_ {\mu\nu}(x,k)}{\partial k_\alpha}\frac{\partial \chi^\gamma}{\partial x^\alpha}k_\gamma + g_{\alpha\nu}(x,k)\frac{\partial\chi^\alpha}{\partial x^\mu}+ g_{\alpha\mu}(x,k)\frac{\partial\chi^\alpha}{\partial x^\nu}\,=\,0\,,
\label{eq:killing}
\end{equation}
where $\chi^\alpha=\chi^\alpha(x)$ is momentum independent. In GR, where the metric does not depend on the momentum, the previous condition reduces to the standard form
\begin{equation}
\chi_{\mu\,;\nu}+\chi_{\nu\,;\mu}\,=\,0\,.
\label{eq:killing_GR}
\end{equation}
Noticeably, it was shown in~\cite{Relancio:2020rys} that promoting the metric to the cotangent bundle does not affect the space-time isometries, therefore preserving the Killing vectors $\chi^\alpha$ of the GR case.   This point will be crucial in the following, as it implies that the whole causal structure of the base spacetime is preserved. 

\subsection{Relativistic deformed  kinematics in curved spacetimes}
%We summarize here our previous results about supplementing a relativistic deformed  kinematics within a curved space-time. 

The deformed kinematics of DSR are usually obtained from Hopf algebras~\cite{Majid:1995qg}, with the $\kappa$-Poincaré kinematics~\cite{Majid1994} being the most studied example of this kind of construction. In~\cite{Carmona:2019fwf}, the deformed DSR kinematics has been given a fully geometric interpretation. Given a de Sitter momentum metric $ \bar{g}$, translations can be used to define the associative deformed composition law, the Lorentz isometries lead to the Lorentz transformations, and the (square of the) distance in momentum space is identified with the deformed Casimir.   

In~\cite{Relancio:2020zok,Pfeifer:2021tas}, some of the present authors extended~\cite{Carmona:2019fwf} in order to consider a deformed kinematics and a curved spacetime within the same framework. To that aim, it is mandatory to consider the cotangent bundle geometry discussed above (Eq.~\eqref{eq:line_element_ps}). The cotangent-bundle metric tensor $g_{\mu\nu}(x,k)$, depending on space-time coordinates, can be constructed in terms of the space-time tetrad field, and the metric of the momentum space, $\bar{g}$, by~\cite{Relancio:2020zok, Pfeifer:2021tas}
\begin{equation}
g_{\mu\nu}(x,k)\,=\,e^\alpha_\mu(x) \bar{g}_{\alpha\beta}(\bar{k})e^\beta_\nu(x)\,,
\label{eq:definition_metric_cotangent}
\end{equation}
where $\bar{k}_\alpha=e^\nu_\alpha (x) k_\nu$, and $e^\alpha_\mu(x)$ denotes the tetrad of spacetime. As discussed in~\cite{Relancio:2020zok}, when constructing the metric in this way we assure that the momentum space (fibers in the cotangent bundle) is maximally symmetric if the starting metric $\bar{g}$ also is. Moreover, in~\cite{Pfeifer:2021tas} it was rederived this construction by lifting the symmetries for flat spacetimes to the curved case. 

In~\cite{Relancio:2020rys} it was also proven that the Casimir associated to a given metric (leading to the same trajectories obtained from the solutions of the geodesic equations of the metric)  can be identified with the square of the minimal geometric distance of a momentum $k$ from the origin of momentum space, measured by the momentum space length as induced by the metric.  This relates  the Casimir and the metric in the following way~\cite{Relancio:2020zok}
\begin{equation}
\mathcal{C}(x,k)\,=\,\frac{1}{4} \frac{\partial \mathcal{C}(x,k)}{\partial k_\mu} g_{\mu\nu} (x,k) \frac{\partial \mathcal{C}(x,k)}{\partial k_\nu}\,.
\label{eq:casimir_metric}
\end{equation} 
But, as discussed in~\cite{Carmona:2019fwf}, any function of such a Casimir is also a valid Casimir, which corresponds to a redefinition of the mass (for the massless case they are the same).

\begin{comment}
A very important relation that the Casimir satisfies is that its delta derivative~\eqref{eq:delta_derivative} is zero, i.e.
\begin{equation}
\frac{\delta \mathcal{C}(x,k)}{\delta x^\mu}\,=\,0\,.
\label{eq:casimir_delta}
\end{equation} 
This is a necessary condition derived from the fact the Hamilton equations of motions are horizontal curves~\cite{Relancio:2020rys}, since the covariant derivative of the Hamiltonian vanishes. {({\bf why?})}
\end{comment}

In \cite{Pfeifer:2021tas} it was shown that the most general form of the metric, in which the construction of a deformed kinematics in a curved space-time background is allowed, is  a momentum space metric whose Lorentz isometries are linear transformations in momenta, i.e., a metric of the form
\begin{equation}
\bar g_{\alpha\beta}(k)\,=\,\eta_{\alpha\beta} f_1 (k^2)+\frac{k_\alpha k_\beta}{\Lambda^2} f_2(k^2)\,,
\label{eq:lorentz_metric}
\end{equation}
where $\Lambda$ is the high-energy scale parametrizing the momentum deformation of the metric and kinematics, and we use $\eta=\mathrm{diag}(1,-1,-1,-1)$. From Eq.~\eqref{eq:definition_metric_cotangent}, one obtains the following metric in the cotangent bundle when a curvature in spacetime is present
\begin{equation}
g_{\mu \nu}(x,k)\,=\,a_{\mu \nu}(x)f_1 (\bar{k}^2)+\frac{k_\mu  k_\nu}{\Lambda^2} f_2(\bar{k}^2)\,,
\label{eq:lorentz_metric_curved}
\end{equation}
where $a_{\mu \nu}(x) = e^\alpha_\mu(x) \eta_{\alpha\beta} e^\beta_\nu(x)$ is the curved space-time metric. 

Noticeably, one can then use the definition~\eqref{eq:affine_connection_st} to show that the space-time affine connection is, for the above class of metrics, momentum independent, giving as a result the same affine connection $ \Gamma^\rho_{\mu\nu}(x)$ of GR \cite{Pfeifer:2021tas}. A relevant consequence  of this fact is that for metrics of the form~\eqref{eq:lorentz_metric_curved}, the Raychaudhuri equation for massless particles is also unmodified~\cite{Relancio:2020rys} 
\begin{equation}
  \frac{d\theta}{d\lambda} \,=\,-\frac{1}{2} \theta^2+ \omega_{\alpha \beta}\omega^{\alpha \beta}- \sigma_{\alpha \beta}\sigma^{\alpha \beta}-R_{\alpha \beta} l^\alpha l^\beta\,,
  \label{eq:ray_eq}
\end{equation}
where $\theta$ is the expansion, $\omega$ the torsion, $\sigma$ the shear, and $l^\alpha$ the null vector describing the geodesic congruence. 
This is due to the fact that the Ricci tensor is \emph{momentum independent}, since the connection is, in this case, the GR one. Also, this result can be easily understood from the fact that the momentum metric~\eqref{eq:lorentz_metric_curved} is covariant under linear Lorentz transformations. As a consequence, the dispersion relation is unmodified for the massless case and so it is the propagation of massless particles with respect to the GR case.

The above conclusion could seem in contradiction with the result of~\cite{Cianfrani:2014fia}, where it was shown that the geodesic deviation equation is generically modified in a momentum dependent spacetime. However, this is not the case because our result is strictly related to a very peculiar form of the chosen metric. Indeed, the formalism adopted in~\cite{Cianfrani:2014fia} should lead to our same conclusion once the momentum metric is specialized to~\eqref{eq:lorentz_metric_curved} and the particle is taken to be massless. 

The requirement for the the momentum metric~\eqref{eq:lorentz_metric} to be a de Sitter space, so allowing us to define a relativistic deformed kinematics, implies a relationship between the functions $f_1$ and $f_2$. In particular, in~\cite{Relancio:2020rys}, it was found that in order to conserve the Einstein tensor defined in~\cite{miron2001geometry} --- which has formally the same expression of GR --- one has to impose that the metric~\eqref{eq:lorentz_metric} is conformally flat. Then, taking $f_2=0$ and imposing a de Sitter momentum space, one obtains from Eq.~\eqref{eq:lorentz_metric_curved} 
\begin{equation}
g_{\mu\nu}(x,k)\,=\,a_{\mu \nu}(x)\left(1-\frac{\bar{k}^2}{4\Lambda^2}\right)^2\,.
\label{eq:conformal_metric}
\end{equation}

This metric is characterized by remarkable properties. In particular, the basis corresponding to Eq.~\eqref{eq:conformal_metric} is the unique one in which the Killing equation~\eqref{eq:killing} can be written equivalently to GR as in Eq.~\eqref{eq:killing_GR}.  As a consequence of this, it can be shown~\cite{Relancio:2021asx} that only for the metric~\eqref{eq:conformal_metric}, the commonly used notions of surface gravity for a Killing horizon~\cite{Cropp:2013zxi} coincide as in GR. 

\begin{comment}
{\sf Note however, that for the extension of the space-time thermodynamics framework discussed in this work, we do not need to restrict {\em a priori} the cotangent bundle geometry to the form~\eqref{eq:conformal_metric}, given that the more general form~\eqref{eq:lorentz_metric_curved} is sufficient for assuring that at least a  subset of the surface gravity definitions are equivalent~\cite{Relancio:2021asx,Relancio:2022kpf}. } 
\red {I think we should ditch this comment given that in the end we do have to specialize to the form~\eqref{eq:lorentz_metric_curved} and moreover I do not like to talk about surface gravity given that $\kappa$ will be only a book keeping parameter that can be rescaled to one. SL}
\end{comment}
%\textcolor{red}{{\bf Better now?}} {Not yet in a way [G]. I slightly changed the sentence; I guess you refer to some literature like {\bf arXiv:1302.2383v1 [gr-qc] 11 Feb 2013}, Am I right? However, what is the relevance of such recovered degeneracy in our setting? We should stress it as it seams to be a key argument for univocally defining the local Rindler horizon temperature in the following. Even before facing the issue of a missing QFT in the DSR  framework.}

\section{Accelerated observers in a momentum dependent geometry}
\label{sec:rindler}

{ We are interested in describing the physics of an accelerated observer with a metric of the form of Eq.~\eqref{eq:lorentz_metric_curved}. We start by deriving the motion of such observers. Thereby, we use a change of coordinates for writing the trajectories in (some deformed) Rindler coordinates and studying a local Rindler wedge description, compatibly with our construction of a cotangent bundle metric~\eqref{eq:lorentz_metric_curved}.}

\subsection{Derivation of trajectories}
 First, we start by noticing from Eq.~\eqref{eq:casimir_metric} that, since the metric~\eqref{eq:lorentz_metric} is invariant under the usual Lorentz transformations, the Casimir must be a function of the squared momentum. Therefore, by inverting this relation one can write 
  \begin{equation}
k^2\,=\, h(m^2/\Lambda^2)\,,
\label{eq:momentum_squared}
\end{equation}
 being $ h(m^2/\Lambda^2)$ some function satisfying $\lim_{\Lambda \to \infty} h= m^2$.  This also means that there is a simple relation between the velocities $v^\nu$ and the momenta, which can be derived as usual by taking the derivative of the Casimir with respect to the momentum, so to obtain
 \begin{equation}
k_\mu\,=\,  \sqrt{h} \eta_{\mu \nu}v^\nu \,.
\label{eq:momentum_velocity}
\end{equation}
%where $h_2=h/m^2$. {\bf Do we really need to introduce $h_2$ given its trivial relation with $h$ ?} 

Keeping this in mind, we start by computing the $\gamma$ factor relating the proper time and the temporal coordinate with the metric~\eqref{eq:lorentz_metric_curved}, in the limit  $a_{\mu\nu} \to \eta_{\mu\nu}$ and $k_\alpha \to \bar k_\mu = k_\mu$, following the prescription used in~\cite{Relancio:2020mpa}.
%{For whatwe discussed last friday, shouldn't we use \eqref{eq:lorentz_metric_curved}? Better, we should probably at the beginning of this section introduce the idea of  a local patch in the general curved and momentum dependent metric; then restrict to the spacetime locally flat case, and only then consider a uniformly accelerated observer. This would also explain why next \eqref{eq:lorentz_metric_mass_cov} does not depend on $x$ in the limit $a_{\mu\nu} \to \eta_{\mu\nu}$.} 
We can write the line element for spacetime as 
 \begin{equation}
d\tau^2\,=\, dx^\mu  \bar{g}_{\mu\nu}(k)dx^\nu\,.
\label{eq:lorentz_metric_mass_cov}
\end{equation}
%{\bf Note: I added a bar for the momentum space metric above to be consistent with the previous notation}

Dividing the previous expression by $dt^2$, and using Eqs.~\eqref{eq:lorentz_metric} and~\eqref{eq:momentum_velocity}, we find (for the sake of simplicity we will work in 1+1 dimensions)
 \begin{equation}
 \left(\frac{d\tau}{dt}\right)^2\,=\, \bar f_1 (m^2)\left(1-v^2\right)+ \bar f_2 (m^2) \left(\frac{d\tau}{dt}\right)^2  \,,
 \label{eq:barf}
\end{equation}
where we have used Eq.~\eqref{eq:momentum_squared}, and defined $\bar f_1 $ and $\bar f_2$, which are some functions involving $f_1$, $f_2$,  $h$ and $m^2$. 
%{ {``some functions'' sounds a bit obscure. We can put the whole calculation at this stage, especially for the second term on the rhs, where we contract momenta and velocities} }.
From the previous equation, we get
\begin{equation}
\gamma \,=\,  \frac{dt}{d\tau} \,=\, \sqrt{ \frac{1-\bar f_2  }{\bar f_1  \left(1-v^2\right)}} \,.
\label{eq:lorentz_metric_mass}
\end{equation}

An  accelerated observer is characterized by a four acceleration vector
 \begin{equation}
\alpha^\mu\,=\,(\alpha^0,\alpha)\,=\, \frac{du^\mu}{d\tau}\,=\,\gamma\frac{du^\mu}{dt }\,=\,\gamma\left(\frac{d \gamma}{dt },\frac{d (\gamma v)}{dt }\right) \,.
\label{eq:4_acceleration}
\end{equation}
We are interested in uniformly accelerated observers for which the spatial proper acceleration can be found by going to their instantaneous rest frame, i.e., setting  $v=0$, and correspondingly,  $\gamma=1$ and $d\gamma/dt=0$. Therefore, from~\eqref{eq:4_acceleration} we can set 
 \begin{equation}
\alpha\,=\, \frac{d (\gamma v)}{dt } \,=\, \text{const}\,.
\label{eq:acceleration}
\end{equation}
By integrating this expression, and assuming $v(t=0)=0$, we are able to express the velocity as a function of the acceleration and time
 \begin{equation}
v\,=\, \frac{\sqrt{ \bar f_1}\alpha  t }{\sqrt{1-\bar f_2 + \bar f_1 \alpha ^2 t^2}} \,.
\label{eq:velocity}
\end{equation}
Integrating this expression in time we can find the position
 \begin{equation}
x\,=\,\frac{\sqrt{1-\bar f_2 +\bar f_1 \alpha ^2 t^2} }{\sqrt{ \bar f_1}\alpha } \,.
\label{eq:x_t}
\end{equation}
By squaring the above expression, {we get the worldline of an observer in hyperbolic motion having constant proper acceleration $\alpha$ in the $x$-direction, that is}
 \begin{equation}
x^2 -t^2\,=\,\ell^2 \,,
\label{eq:hyperbola}
\end{equation}
with 
\begin{equation}
 \ell\,=\, \frac{\sqrt{1- \bar f_2  }}{\sqrt{ \bar f_1} \alpha}\,.
\label{eq:chi}
\end{equation}
{It is convenient to set $\ell = \mu/\alpha$, with $\mu \equiv \sqrt{1- \bar f_2  }/\sqrt{\bar f_1}$ carrying the effect of the momentum dependent geometry. 
 We can see, e.g~from Eq.~\eqref{eq:barf}, that the factor $\mu$ depends only on the mass of the particle probing the spacetime, not on its momentum. Therefore, the worldline in \eqref{eq:x_t} defines a family of hyperbolae, where $\chi$ varies for observers with same uniform acceleration and different mass.}

In the limit $\Lambda$ going to infinity, $\bar f_1 \to 1$ and $\bar f_2 \to 0$, so $\mu \to 1$, and the usual result of GR is obtained.

\subsection{Change to Rindler coordinates}
{Along with \eqref{eq:hyperbola}, the worldline of the uniformly accelerated observer can be written
in hyperbolic (polar) coordinates as
 \begin{equation}
t\,=\, \ell \sinh\left(\kappa\eta \right)\,,\qquad x\,=\, \ell \cosh\left(\kappa \eta \right)\,,
\label{eq:rindler_change}
\end{equation}
%with inverse
 %\begin{equation}
%\eta\,=\, \int \frac{1}{\gamma}\, dt\,=\,  \text{arctan}\left(\frac{\sqrt{ \bar f_1}\alpha  t }{\sqrt{1- \bar f_2   + \bar f_1 \alpha ^2 t^2}}\right) \,, \qquad \chi=\sqrt{x^2-t^2}\, ,
%\label{eq:tau}
%\end{equation}
% \blue{\begin{equation}
%\eta\,=\, \int \frac{1}{\gamma}\, dt\,=\,  \text{arctan}\left(\frac{ \alpha  t }{\sqrt{\mu   +  \alpha ^2 t^2}}\right) \,, \qquad \chi=\sqrt{x^2-t^2}\, ,
%\label{eq:tau}
%\end{equation}}
%providing the transformation formulas between the inertial and the hyperbolic coordinates. 
where $\eta$ is the hyperbolic angle, and $\kappa$ is an arbitrary bookkeeping parameter with the dimensions of an acceleration (which can always be set to one by a suitable rescaling of the proper time on a given accelerated observer worldline). Given that $\mu >0$ (which is always the case for on-shell particles, since $m \ll \Lambda$), the inertial coordinates cover the so-called Rindler wedge region $0<x<\infty ,\;x>|t|$.}
%On the hyperbola located at $\chi = \chi_0$ (see Figure \ref{}) the proper time of the uniformly accelerated observer is $d\tau = \chi_0 d\eta$. Therefore, we can generally express the hyperbolic angle as a function of proper time as 
%\begin{equation}
%\eta= \chi^{-1}\, \tau= \frac{\alpha\, %\tau}{\mu}\, .
%\label{h_angle}    
%\end{equation}}

By taking the metric in \eqref{eq:lorentz_metric_curved}, in the locally flat limit, we have 
\begin{align}
ds^2\,=\, \left(f_1 + f_2 \frac{k_0^2}{\Lambda^2}\right) dt^2+2  f_2 \frac{k_0 k_1 }{\Lambda^2} dt\, dx - \left(f_1- f_2 \frac{k_1^2}{\Lambda^2}\right)dx^2\,.
\end{align}
Thereby, via the change of coordinates in \eqref{eq:rindler_change},
% \begin{equation}
%t \,=\,   \chi  \sinh{(\alpha  \tau)}\,,\qquad x\,=\,  \chi   \cosh{(\alpha  \tau)}\,,
%\label{eq:rindler_change}
%\end{equation}
%satisfying the hyperbolic motion~\eqref{eq:hyperbola}, where we have made the identification
%\begin{equation}
%\chi\,=\, \frac{\sqrt{1- \bar f_2  }}{\sqrt{ %\bar f_1} \alpha }\,.
%\label{eq:chi}
%\end{equation}
we get the deformed Rindler line element
\begin{align}
ds^2 \,=\, \kappa^2\left(f_1 \ell^2 + f_2 \frac{k_0^{\prime 2}}{\Lambda^2}\right) d\eta^2+2 \kappa f_2 \frac{k^\prime_0 k^\prime_1 }{\Lambda^2} d\eta d\ell + \left(-f_1+ f_2 \frac{k_1^{\prime 2}}{\Lambda^2}\right)d\ell^2\,, 
%ds^2\,&=\, \left(f_1 \chi^2 \frac{\alpha^2}{\mu^2} + f_2 \frac{k_0^2}{\Lambda^2}\right) d\tau^2 +2  f_2 \frac{k_0 k_1 }{\Lambda^2} d\tau\, d\chi - \left(f_1- f_2 \frac{k_1^2}{\Lambda^2}\right)d\chi^2 \, , 
\label{eq:rindler_ds2}
\end{align}
with
 \begin{equation}
g_{00} \,=\, \kappa^2\left( f_1 \ell^2  + f_2 \frac{k_0^{\prime 2}}{\Lambda^2}\right)\,,\qquad 
g_{01} \,=\,\kappa f_2 \frac{k^\prime_0 k^\prime_1}{\Lambda^2}\,,\qquad 
g_{11} \,=\,- f_1+ f_2 \frac{k_1^{\prime 2}}{\Lambda^2}\,.
\label{eq:rindler_metric_easy}
\end{equation}
Differently from the standard case, to get Eq.~\eqref{eq:rindler_metric_easy} here we need to take into account that also momenta change as coordinates change. Indeed, $k_\mu dx^\mu$ is invariant under a coordinate change of the form $x^{\prime \mu}  =x^{\prime \mu}(x)$,
and $k^{\prime}_\mu=\frac{\partial x^\nu}{\partial x^{\prime\mu}} k_\nu$, $k$ and $x$ being canonically conjugate variables. Because of that, and due to the fact that $k^2$ is invariant too,  as $k_\mu \eta^{\mu \nu}k_\nu  =k^\prime_\mu a^{\mu \nu} k^\prime_\nu$, we get the expression in \eqref{eq:rindler_metric_easy}. To ease the notation, in the following we will use $k$ instead of $k^\prime$ to denote the momenta associated to the Rindler coordinates.

As a final sanity check, let us note that by taking the limits $  f_1 \to 1$ and $  f_2 \to 0$, one recovers the usual Rindler metric in polar coordinates
 \begin{equation}
g_{00} \,=\, \kappa^2 \ell^2\,,\qquad 
g_{01} \,=\, 0\,,\qquad 
g_{11} \,=\,-1\,,
\label{eq:rindler_metric_GR}
\end{equation}
where $\ell$ would be $1/\alpha$.

Some remarks are in order at this stage.
Consistently with the assumption of {linear} Lorentz covariance of the metric, we see that a Lorentz boost still generates a proper-time translation along the uniformly accelerated observer worldline in \eqref{eq:hyperbola}. In GR, by changing from inertial Minkowski $(x,t)$ to hyperbolic coordinates $(\ell, \eta)$, one sees that the boost symmetry is just a symmetry of rotation of the hyperbolic angle $\eta$, i.e., $\chi_B=\partial_{\eta}$. 

In this framework, on the observer hyperbola located at $\ell=\ell_0$, the relation between the hyperbolic angle and the proper time is given by the Rindler line element. We have 
\begin{equation}
 d\tau^2 \,=\, \kappa^2\,\left(f_1 \ell_0^2+ f_2 \frac{ k_0^2}{\Lambda^2}\right) d\eta^2\,\equiv\,\kappa^2\, {\tilde{\ell}_0}^2 d\eta^2\,.
 \label{eq:l_definition}
\end{equation}

We see how the scaling of
the Killing field $\partial \tau = (1/\tilde{\ell}_0)\, \partial_{\eta}$, generating proper time flow on this particular
hyperbola, gets modified in the momentum deformed metric. The difference with respect to the standard Lorentz hyperbola in the Rindler wedge lies in the scaling factor 
\begin{equation}
\tilde{\ell}_0\,=\, \ell_0\,  \sqrt{f_1 + f_2 \frac{ h}{\Lambda^2}} \, ,
    \label{eq:scale}
\end{equation} 
which introduces a dependence on the mass of the accelerated observer (or more precisely, of the detector coupled to the field in order to observe an Unruh temperature): indeed $\ell_0$ is given by Eq.~\eqref{eq:chi}, and $f_1$ and $f_2$ depend on the squared momentum, which is related to some function of the squared mass as in Eq.~\eqref{eq:momentum_squared}.  This allows us to interpret $1/\tilde{\ell}_0$ as a corrected uniform acceleration of the hyperbolic worldline.

Further, we shall relate $k_0$ with the mass of the particle from the dispersion relation~\eqref{eq:momentum_squared}, as written in the Rindler metric. We get 
\begin{equation}
\mathcal C(k)\,=\, k_\mu a^{\mu \nu}k_\nu\,=\, \frac{k_0^2}{\kappa^2\ell^2}-k_1^2\,=\,h\,.
\end{equation}
From the above dispersion relation, we obtain the following equations of motion
\begin{equation}
 \dot x^\mu\,=\,\frac{1}{2} \frac{\partial \mathcal C(k)}{\partial k_\mu} \,.
\end{equation}
As we are fixing $\ell =\ell_0= \text{const}$, $\dot \ell_0=k_1=0$, so we get $k_0^2=h \kappa^2 \ell^2_0$, since the dispersion relation must hold.

%The proper time $\tau$ along the orbit is proportional to the boost parameter $ \tau= \ell_0\, \eta $, hence the Lorentz boost $L(\eta)$ in the $x$ direction  generates a proper-time translation of proper time along the accelerated worldline. The proper time flow is generated by the Killing field $\partial_{\tau} = (1/\ell_0)\, \partial_{\eta}$.

%The modification of the trajectories is of the order of $m^2/\Lambda^2$, where $m$ is the mass of the particle probing the momentum dependent spacetime. Therefore, the modification is very small for elementary particles. }

Most importantly, from Eq.~\eqref{eq:rindler_change} we see from the quotient $\lim_{\tau \to \infty} x/t$ that the Rindler (acceleration) horizon of the wedge coincides with the Killing horizon defined by the set $|x|=|t|$, exactly as in GR. Therefore, the causal structure of the Rindler wedge is not affected by the deformation of the metric within the locally flat patch approximation.

%On the other hand, the distance from the origin to a point $x$ at $t=0$ is not given by $1/\alpha$ as in GR, but it comes with a small modification, viz.,   $ \sqrt{1- \bar f_2 }/\sqrt{\bar f_1} \alpha$. This means that observers with different masses follow different trajectories, despite of suffering the same acceleration. Also, this small modification can be understood as a small correction of the proper acceleration, where a term depending on the mass arises. Again, this correction is extremely small for elementary particles. }

%In this previous subsection we have derived the Rindler metric starting from a momentum dependent spacetime and seeing what is the metric associated to an accelerated observer. However, the same derivation can be done starting from the momentum metric for flat spacetime~\eqref{eq:lorentz_metric_curved}, obtaining the same result. This serves us as a consistency check for our framework. Note that the computations of the previous subsection is mandatory in order to obtain the relationship~\eqref{eq:chi}.{{\bf I think we can remove this paragraph.} } 

\subsection{Rindler horizon and Unruh temperature}

Despite a complete QFT based on the symmetries of DSR is still missing, in the following we shall assume that such framework exists in the specific setting given by \eqref{eq:lorentz_metric_curved}, within the locally flat limit $a_{\mu\nu} \to \eta_{\mu\nu}$ and $k_\alpha \to \bar k_\mu = k_\mu$. Moreover, we will assume that in this still unknown framework that the Bisognano--Wichmann theorem~\cite{Jacobson:2012ei,Martinetti:2002sz} still holds as in GR. For that, we need this theory to be invariant under linear Lorentz and $CPT$ transformations. The first part is obvious from our particular choice of momentum coordinates, so Poincaré group is not deformed. With respect to the latter point, there is no consensus in the present literature given that while some works claim that $CPT$ invariant formulations of DSR field theory are possible~\cite{Carmona:2021pxw,Franchino-Vinas:2022fkh}, there are others reaching opposite conclusions~\cite{Arzano:2020jro,Bevilacqua:2022fbz}. In any case, let us stress that with our choice of the metric~\eqref{eq:lorentz_metric_curved} the equation of motion of massless particles are unmodified and hence, by construction, $CPT$ invariant. 

Another important remark can be made comparing this assumption with the result of~\cite{Kowalski-Glikman:2009yku}, where it was found that the Unruh temperature associated with Rindler wedge leads to a not thermal spectrum. However, this result was obtained for a particular basis of $\kappa$-Poincaré which does not present a linear Lorentz invariance, in contrast with our case of study.

Let us then consider a quantum field theory defined on the base spacetime $M$ endowed with a momentum dependent metric. In the previous section, we showed that the worldline of a uniformly accelerated observer in the deformed Rindler wedge $W$ is still an orbit of the action of a one-parameter group (the Lorentz boost in the $x$ direction). Once parametrized by proper time, the flow in the hyperbolic angle can be interpreted as the time flow of the accelerated observer, with the scaling factor $\tilde{\ell}_0$ in \eqref{eq:scale} playing the role of an effective acceleration. Moreover,  we discussed at the end of the previous subsection that a Rindler (Killing) horizon structure for the deformed Minkowski metric is preserved. 

These facts provide together a compelling argument supporting the existence also in this setting of a 
%Therefore, we hypothesize that the Fulling--Davies--Unruh effect survives the deformations induced by the momentum dependence of the metric in this setting.
%We shall then be able to define a correspondent 
Unruh--Davies temperature for an uniformly accelerated observer in the the deformed Rindler wedge $W$ associated to the  metric~\eqref{eq:rindler_metric_easy}. As the Rindler wedge causal structure is preserved, we still have that each observer instantaneous Cauchy surface is cut into two parts by the edge of the wedge. The vacuum $\Omega$ has correlations across the edge, hence the two sets of degrees of freedom on the two sides of the edge are entangled. 

The effect of the mass dependent deformation of the metric is  to scale the acceleration. Thence, it modifies the proper distance of the observer worldline from the Rindler edge (bifurcation surface). This may introduce a new scale, but it does not affect the structure of correlations of the vacuum.

As a consequence of the entanglement, the observer who only has access to the restriction of $\Omega$ to $W$ sees the vacuum as a thermal mixed state. From the Bisognano--Wichmann theorem~\cite{Jacobson:2012ei,Martinetti:2002sz}, we know that the restriction of the QFT vacuum $\Omega$ to the Rindler wedge (more precisely, over the algebra of the local observables in $W$) is given by a canonical thermal state with density matrix
\begin{equation}
   \rho_R \propto e^{-2\pi K}\,=\,e^{-\beta_R H_{\eta}}\,, 
   \label{eq:thermal_R}
\end{equation}
where $K$ denotes the boost generator, $H_{\eta}= K\,\hbar $ is the so-called boost Hamiltonian, and $\beta_R= 2\pi/\hbar$ is the (inverse of the) Rindler temperature associated to the thermal state.

Written in terms of the accelerated observer’s quantities, the density matrix in \eqref{eq:thermal_R} reads
\begin{equation}
\rho_R \, = \,e^{-\beta_U H_{\tau}}\,, 
\end{equation}
where the operator $H_{\tau} = \hbar K / \kappa \tilde{\ell}_0$ is the boost Hamiltonian scaled to generate translations of
this proper time, while the corresponding (inverse)
temperature is given by the Unruh--Davies temperature 
\begin{equation}
    \beta_U \,=\, \frac{2\pi\, \kappa\, \tilde{\ell}_0}{\hbar}\, .
    \label{eq:u_temp}
\end{equation}

In particular, using~\eqref{eq:scale} and~\eqref{eq:chi} one can see how the Unruh temperature gets modified in our setting with respect to GR 
\begin{equation}
   T_U^{\Lambda} \,=\, \frac{\hbar \alpha}{2\pi } \frac{\sqrt{\bar f_1}}{\sqrt{\bar f_1+\bar f_2}\sqrt{1-\bar f_2}}\,.
   \label{eq:temperature_Lambda}
\end{equation}
The momentum induced deformation of the metric therefore reflects in a modification of the Unruh temperature\footnote{Note however, that for the particular metric~\eqref{eq:conformal_metric} the temperature does not change with respect to the GR result. }, while the usual GR expression is recovered in the limit $\Lambda \to \infty$
\begin{equation}
  T_U^{\Lambda\to \infty}\, = \,T_U^{GR} \,=\, \frac{\hbar \alpha}{2\pi }   \,.
   \label{eq:temperature_GR}
\end{equation}

Accordingly, we can get Rindler Wedge $T_R=1/\beta_R$ from the Unruh--Davies temperature by rescaling via the momentum dependent gravitational Doppler factor, that is
\begin{equation}
T_R \approx T_U \, \sqrt{g_{00}}= \frac{\hbar\, \kappa}{2\pi}\, .
     \label{R_temperature}
\end{equation}
The Rindler temperature $T_R$ stays constant throughout the  wedge and it is well defined on the horizon. Therefore, the thermal character of the Rindler state is effectively extended from the single Rindler observer to the whole wedge.

\section{Einstein's equations from space-time thermodynamics}
\label{sec:einstein}

The thermal character of the vacuum of a relativistic quantum field theory in the Rindler wedge allows to think of $W$  as a (local) thermal system bounded by Rindler Killing horizon membrane. The thermal behavior of the system is due to the entanglement between the quantum states in the right and left wedges. The very same entanglement can be accounted for the entropy of the thermal wedge, measured via the von Neumann entropy of the state $\rho_R$, and shown to be proportional to the area of the Rindler horizon bifurcation surface
\begin{equation}\label{AS}
    S\,=\, \zeta A\,,
\end{equation}
with proportionality factor $\zeta$ a priori depending on the nature of the quantum fields, as well as being some complicate function of the position in spacetime. In our setting, this factor can also depend on the momentum. 
In the remarkable work~\cite{Jacobson:1995ab}, the author showed how the thermal equilibrium character of the quantum field system at local level is essentially intertwined with the Killing character of the Rindler horizon structure, namely a on-shell space-time geometric configuration with respect to GR locally \footnote{The essentially ultra-local and geometric nature of the thermal behavior of the Rindler horizon finds support down to a full quantum gravity level of description, as more recently shown in \cite{PhysRevD.90.044044, Chirco_2015}. See also \cite{Chirco:2015bps}.}. In the same work, this relation was inverted to show that the very Einstein equation could be derived as a space-time thermodynamics equation of state, starting from a \emph{local} constitutive equilibrium equation for the space-time degrees of freedom. 

In fact, given a space-time patch around any point $p$ in $M$, a local Rindler horizon structure can be introduced via local Lorentz invariance in the \emph{local} inertial frame in $p$ by a Lorentz boost to an uniformly accelerating frame. Accordingly, once reduced to the local Rindler wedge $R$, the locally Minkowski vacuum gets described by a thermal state $\rho_R$, with the local Rindler horizon playing the role of a diathermic membrane. For a small perturbation to the thermal vacuum $\delta \rho_R$, the thermodynamic Clausius relation between a change of entropy and an energy flux across the local Rindler horizon holds,
\begin{equation}\label{clausius}
T\, dS\,=\,\delta Q \,,
\end{equation}
 leading, via \eqref{AS}, to a constitutive relation connecting changes in the local wedge geometry, expressed in terms of horizon area deformations $\delta A$, to fluxes of matter's stress-energy tensor currents intended as heat flux through the horizon.
 %\begin{equation}\label{flux}
%\delta Q\,= \,  \int_H T_{\mu \nu}\chi^\mu_B d\Sigma^\nu \,,
%\end{equation}
%with $d\Sigma^\nu$ the horizon volume element. Note that from now on we shall drop the $B$ label for the boost Killing vector to streamline our notation.
 
In the following, we shall reconsider the original derivation in~\cite{Jacobson:1995ab} for our deformed space-time setting, to investigate whether the thermodynamic characterization of Einstein's equations is affected by the momentum dependence of the DSR metric. 

\subsection{Local Rindler horizon}

\begin{figure}
\centering
\includegraphics[width=8cm]{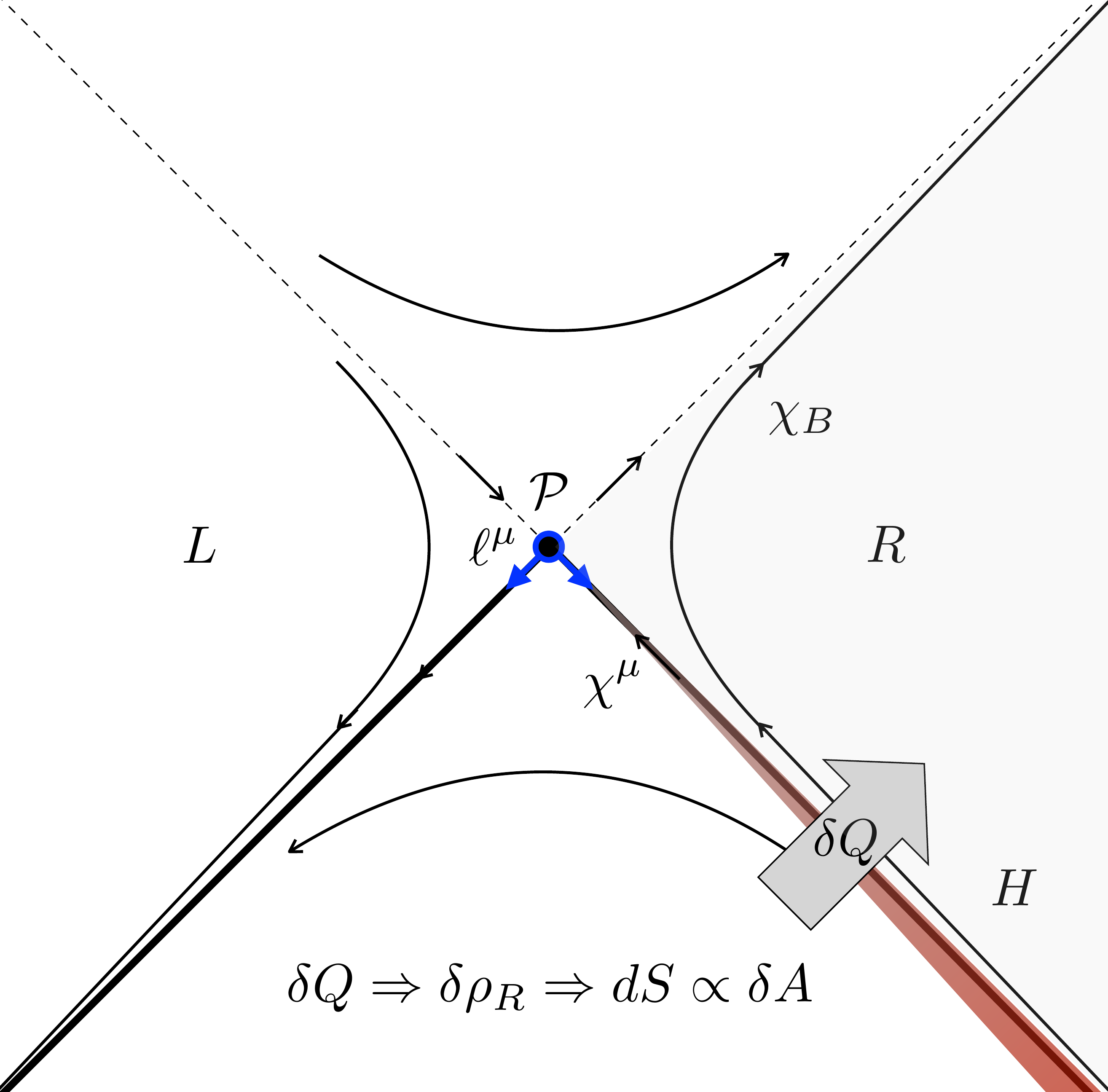}
\caption{Local Rindler horizon framework. The oriented hyperbola indicates the approximate boost Killing flow $\chi_B$ in locally Minkowski spacetime. Given a generic point $p$ in $M$, a local horizon at $p$ is defined as one side of the boundary of the past of a spacelike 2-surface patch $\mathcal{P}$ including $p$. The local horizon is comprised by the congruence of null geodesics characterized by the past pointing tangent null vector $\ell^\mu$ (in blue). Such a structure coincides with a local Rindler horizon as the local inertial frame in the neighborhood of $p$ is boosted to a uniformly accelerated frame. The approximate boost field on the local Rindler horizon is denoted by $\chi^\mu$. Assuming the ground state of the fields living in the spacetime to be locally approximated by the Minkowski vacuum, we get a local description of the thermal Rindler wedge as a consequence of the thermal character of the reduced vacuum on the right wedge $(R)$. An infinitesimal perturbation of the fields equilibrium density matrix is associated to a variation of the horizon entropy, which reflects in an infinitesimal perturbation of the horizon geometry.  
%\blue{In this figure the flux $\delta Q$ arrow is horizontal which makes you think of a superluminal (actually infinite) speed flux. Can we make it almost vertical (bottom-up) or at 45 degrees instead?} 
} 
\label{figure_LCH}
\end{figure}

{Let us start by considering the approximate Killing description of a \emph{local} Rindler causal horizon in the cotangent bundle setting. In analogy with the global definition of a horizon as the boundary of the past of future null infinity, one can generally consider a local horizon at \emph{any} $p$, in the \emph{base} spacetime $(M,g_{\mu \nu})$, as one side of the boundary of the past of a spacelike 2-surface patch $\mathcal{P}$ including $p$. Near $p$, the local horizon is comprised by the congruence of null geodesics orthogonal to $\mathcal{P}$, characterized by the past pointing tangent null vector $\ell^{\mu}$. 

In particular, within a small neighborhood of $p$ in $M$, local Lorentz invariance allows to map the local inertial frame in $p$ to a local Rindler frame via Lorentz boost. Then, locally, the boundary of the past of the patch $\mathcal{P}$ gets mapped to a section of an approximate Killing horizon, centered in $p$. The future pointing approximate boost Killing vector $\chi^\mu$ vanishes in $p$ and it is tangent to the null congruence comprising the causal horizon (see Figure \ref{figure_LCH}).}
%Consider any point $p$ in the base space-time $M$ as sitting on a codimension-$2$ space-like hypersurface $\mathcal{P}$, with future-directed null vectors $\ell^\mu$ on $\mathcal{P}$ perpendicular to the surface. The local causal horizon is constructed as the past of these null vectors.
%We write the affine parameter of each null vector as $\lambda$ and we set $\lambda= 0$ on $\mathcal{P}$. {\bf In Jacobson's original derivation, the thermal system coincide with the region $L$ in spacetime behind the local causal horizon as perceived by the uniformly accelerated observer $\chi_{-}$. We choose to reproduce the argument by taking $R$ as the thermal system with respect to the uniformly accelerated observer $\chi$ and focus on the future horizon.}\\
%\red{I am perplexed here: if you look back at Jacobson '95 he takes the left wedge but with inverse Killing flow w.r.t. the one in shown in our figure. So it is basically the same as taking the Right wedge and having the flow coming in from the past horizon.}

Let us now introduce, along the horizon null hypersurface, a time label $\mathrm{v}$ with respect to the approximate Killing field, ${\chi^\alpha}_{;\alpha} \mathrm{v}=1 $. Further, we express this Killing parameter $\mathrm{v}$ in terms of the null congruence affine parameter $\lambda$. For a Killing horizon, the relation is generically given by 
\begin{equation}
   \lambda \,=\,- e^{-\kappa \mathrm{v}}\,,
\end{equation}
so that the point $p$ is located at infinite Killing parameter
and at $\lambda = 0$. Note that, for dimensional consistence of the local boost Killing flow, we keep for the Killing parameter $\mathrm{v}$ along the horizon the same scaling $\kappa$ we introduced for the hyperbolic angle $\eta$ defined in the wedge.

As a consequence, one gets $\chi^\mu = (d\lambda/d \mathrm{v})\, \ell^\mu$, with
$(d\lambda/d \mathrm{v})=-\kappa \lambda$, which can in turn be used to derive
some helpful relations between the null congruence expansion
$ \hat{\theta}$ and shear  $\hat{\sigma}$ in the Killing parameter, and the respective  quantities in the affine parameter.\footnote{{The twist vanishes instead due to the hypersurface orthogonality of the null congruence of the horizon generators.}}
\begin{equation}
   \hat{\theta}\,=\,\left(\frac{d \lambda}{d \mathrm{v}}\right)\theta\,=\,  -\kappa\, \lambda\, \theta\,,\qquad  \hat{\sigma}\,=\,\left(\frac{d \lambda}{d \mathrm{v}}\right)\sigma\,=\,  -\kappa \,\lambda\, \sigma\,.
\end{equation}

\subsection{Local Rindler wedge perturbation}
{Following the argument in \cite{Jacobson:1995ab} --- and on the base of our previous discussion concerning the thermal state of the Rindler wedge within our framework --- we assume that the ground state of the fields living in the base spacetime is approximated by the Minkowski vacuum. Then, with respect to the approximate Killing vector flow in the Rindler wedge, the vacuum state can be interpreted as an approximate thermal state, with a wedge temperature previously introduced in \eqref{R_temperature} as
\begin{equation}
T_R \approx T_U \, \sqrt{g_{00}}= \frac{\hbar\, \kappa}{2\pi}\, .
\end{equation}
 }
 
%We also write the temperature as a function of the parameter $\kappa$, corresponding to the Rindler horizon surface gravity. {As discussed in the previous section, this is given by the scaling of the Killing field $\partial_{\tau} = (1/\ell_0)\, \partial_{\eta}$ that generates proper-time flow on the hyperbolic observer worldline, namely $\kappa = 1/\ell_0$. We have then
%\begin{equation}
%   T \,=\, \frac{\hbar \kappa}{2\pi}\,= \, \frac{\hbar}{2\pi \ell_0}\,. 
%\end{equation}

%In~\cite{Jacobson:1995ab,Chirco:2009dc} it is discussed how to include an entropy for the Rindler wedge. In particular, by introducing a ultraviolet cutoff one can make this entropy to become proportional to the area of the local boundary of the Rindler wedge, that is
%\begin{equation}
%    S\,=\, \beta A\,.
%\end{equation}
%The proportionality factor $\beta$ can  a priori depend on the nature of the quantum fields as well as be some complicate function of the position in spacetime. In principle, it can also depend on the momentum. 

Further, along with~\cite{Jacobson:1995ab,Chirco:2009dc}, we shall consider an area scaling for the entanglement entropy of the fields reduced density matrix restricted on the right Rindler wedge. This assumption is based on the fact that, as we have seen above, the causal structure of the Rindler wedged is preserved in our framework. This implies that we still have two causally disconnected wedges linked at the bifurcation surface. Therefore, we do expect the entanglement entropy associated to one wedge to  be still proportional to the area of this surface. In particular, 
we shall take $ S\,=\, \zeta\, A$, where $A$ is the are of the bifurcation surface of the Rindler horizon, while we are assuming the proportionality factor $\zeta$ to be a constant. Consequently, we can think of an infinitesimal perturbation of the thermal Rindler system as inducing an entropy change proportional to an infinitesimal horizon area variation, that is
\begin{equation}
   d S\,=\, \zeta\, \delta A\,.
\end{equation}
The variation in the area can be computed from the expansion rate of the congruence of null geodesics comprising the horizon, i.e.
\begin{equation}
   \delta A\,=\, \int_H \tilde{\epsilon} \, \theta \, d \lambda \,,
   \label{eq:area}
\end{equation}
where $H$ indicates we are integrating along the Rindler wedge horizon, and $\tilde{\epsilon}$ is the 2-surface area element of the horizon cross-section. Moving away from the equilibrium bifurcation surface at $\lambda=0$, along the null congruence, the infinitesimal evolution of $\theta$ is given by its linear expansion at the point $p$,
\begin{equation}
   \theta\,\approx\, \theta_p +\lambda \frac{d \theta}{d \lambda}\Bigr\rvert_{p}+\mathcal{O}(\lambda^2) \,.
\end{equation}
The linear coefficient can be determined from the Raychaudhuri equation~\eqref{eq:ray_eq}. 
This means that the entropy can be written from  Eq.~\eqref{eq:area} as
\begin{equation}
   d S\,=\, \zeta  \int_H \tilde{\epsilon}\,  d \lambda \left[\theta -\lambda\left(\frac{1}{2}\theta^2+ ||\sigma||^2 +R_{\mu \nu} \ell^\mu \ell^\nu\right) \right]\,,
    \label{eq:entropy_a}
\end{equation}
where we used the notation $||\sigma||^2=\sigma_{\mu \nu}\sigma^{\mu \nu}$. 

Now we use the thermal equilibrium description for the quantum fields in the Rindler system. For an infinitesimal variation around equilibrium, the entropy change is matched by the amount of heat transferred to the system divided by the equilibrium temperature. This linear response relation corresponds to the Clausius formula in thermodynamics
\begin{equation}
   d S\,=\, \frac{\delta Q}{T_R}\,,
\end{equation}
where the heat variation $\delta Q$ is here expressed as a flux of the quantum field stress-energy tensor $ T_{\mu \nu}$ across the horizon membrane 
\begin{equation}
 \delta Q\,=\,  \int_H T_{\mu \nu}\chi^\mu d\Sigma^\nu \,,
\end{equation}
with $d\Sigma^\nu=\tilde{\epsilon}\, d\lambda \,\ell^\nu$. In our setting, the stress-energy tensor can depend, in principle, also on the momentum, as it is derived from the variation of a Lagrangian density with respect to the deformed metric.

The previous equation, once expressed in terms of the null congruence parameters, reads
\begin{equation}
 \frac{\delta Q}{T_R}\,=\,  \int_H  \tilde{\epsilon}d\lambda \left(-\frac{\lambda }{T_R}\kappa\right) T_{\mu \nu}\ell^\mu \ell^\nu \,.
 \label{eq:entropy_q}
\end{equation}
%At this stage, for $T= \hbar \kappa/2\pi$, we see that the dependence on the particular observer encoded in the  parameter $\kappa$ is eliminated. 

Now, imposing a Clausius relation then amounts to equating the integrands of Eqs.~\eqref{eq:entropy_a} and \eqref{eq:entropy_q}. At zero order in $\lambda$, the heat flux at $p$ is zero, so necessarily $\theta_p=0$. At first order, we get 
\begin{equation}
    \frac{2 \pi}{\hbar \zeta} T_{\mu \nu} \ell^\mu \ell^\nu \,=\, \left(||\sigma||^2 +  R_{\mu \nu} \ell^\mu \ell^\nu\right)_p\,.
\end{equation}
When requiring that $\sigma_p=0$, one finds
\begin{equation}
    \frac{2 \pi}{\hbar \zeta} T_{\mu \nu}  \,=\,  R_{\mu \nu} + \Phi a_{\mu \nu}\,,
    \label{eq:ET}
\end{equation}
where we used that, if $ a_{\mu \nu}(x) \ell^\mu \ell^\nu=0$,   $ g_{\mu \nu}(x,k) \ell^\mu \ell^\nu \neq 0$, since $\ell^\mu \propto \chi^\mu$, so $k_\mu \chi^\mu\neq 0$, being that this is the definition of Killing energy, which is also valid in this scheme~\cite{Relancio:2020rys}.

The next ingredient consists in requiring the conservation of the stress energy tensor,  by which we get 
\begin{equation}
 g^{\rho \mu} \left(R_{\mu \nu}+ \Phi a_{\mu \nu} \right)_{;\rho}\,=\,0\,.
  \label{eq:cov_t}
  \end{equation}

This equation is  quite cumbersome to solve for $\Phi$  since, in this cotangent bundle scenario, the Bianchi identities are not satisfied in general. So we can start by considering a simple example of space-time metric, a Friedmann-Lemaître-Robertson-Walker universe.  When making a power series expansion in $\Lambda$ we find that the only way in which Eq.~\eqref{eq:cov_t} can be satisfied is by imposing that $f_2=0$.

This can be easily understood. For $f_2=0$,  the Bianchi identities can be written exactly as in GR due to the conformally flat form of the momentum metric, since the momentum dependence of the Ricci tensor  cancels out with the corresponding term of the metric.  Explicitly,
\begin{multline}
  g_{\mu \nu}(x,k) R(x,k)\,=\,  g_{\mu \nu}(x,k) g^{\rho \sigma}(x,k) R_{\rho \sigma}(x) =\\
  \,=\, \left(1-\frac{\bar k^2}{4\Lambda^2}\right)^2 a_{\mu\nu}(x)  \left(1-\frac{\bar k^2}{4\Lambda^2}\right)^{-2} a^{\rho \sigma}(x) R_{\rho \sigma}(x) \,=\, a_{\mu\nu}(x)R(x)\,.
\end{multline}
Taking into account all this it is easy to get $\Phi=-1/2 \,R+ \Lambda_0+\varphi (\bar k^2)$. 

The presence of the last term of $\Phi$ is quite obvious, since the covariant derivative of any function of the Casimir is zero, as showed in~\cite{Relancio:2020rys}. Further, $\varphi$ must be such that  $\varphi (0)=0$, since for $\Lambda$ going to infinity we want to recover the usual Einstein equations. Therefore, this must be a small correction, which indeed is proportional to the squared quotient of mass of the particle and the high-energy scale.

Substituting this result into \eqref{eq:ET} we finally find
\begin{equation}
    \frac{2 \pi}{\hbar \zeta} T_{\mu \nu}  \,=\,  R_{\mu \nu} -\frac{1}{2} R\,  g_{\mu \nu}   + (\Lambda_0+\varphi) \, a_{\mu \nu}\,,
    \label{eq:EE}
\end{equation}
where $\Lambda_0$ is some arbitrary integration constant. Thereby, for
\begin{equation}
    \zeta\,=\, \frac{1}{4\hbar G}\,,
\end{equation}
one gets the Einstein equations for the local thermal \emph{base} spacetime description in the cotangent bundle framework, where now both the metric and the Ricci scalar are substituted by their momentum dependent version and a new term, proportional to the squared of the momentum of the particle probing spacetime, appears. {Noticeably, assuming the Equivalence Principle holds in this setting, the above construction can be done at any point in spacetime $p$, hence equation \eqref{eq:EE} must hold everywhere.} 

In~\cite{miron2001geometry}, a prescription for constructing the Einstein equations in spacetime was proposed. The main difference with respect to Eq.~\eqref{eq:EE} is that the definition of the Riemann tensor differs from the one of GR, leading to a different Ricci tensor, and the last term proportional to $\varphi$ is missing. 

In~\cite{Relancio:2020rys}, some of the present authors also discussed the fact that, when considering a cotangent bundle geometry, there are four curvature tensors~\cite{miron2001geometry}: one associated to spacetime, another one to momentum space, and another two mixing momentum space and spacetime. Therefore, one can construct four Einstein's equations including these tensors~\cite{miron2001geometry}.  If the mixing curvature tensors do not vanish, it would require an energy-momentum tensor for such intertwined spaces, which are very difficult to justify from a physical point of view, given the far from clear nature of the sources that would generate such tensors.  Assuming this prescription, one is forced to consider that $f_2=0$, since it is the only metric compatible with vanishing energy-momentum tensor for the intertwined spaces~\cite{Relancio:2020rys}. This is exactly the same result obtained in this paper, even if we considered here a more general form of the Einstein equations. 

For the particular case in which $\varphi=0$,  Einstein's equations are indeed exactly the same of the ones of GR, without any momentum modification of the GR Einstein equations. Moreover, since the right hand side of Eq.~\eqref{eq:EE} does not depend on the momentum, this automatically implies that the energy-momentum tensor can neither depend on it, so the same sources of the GR metrics are also sources of this construction of a cotangent bundle metric.

Alternatively, one can set to zero only the integration constant $\Lambda_0$ and try to make up for the observed dark energy component of our universe via $\varphi$. 
Indeed,  it is easy to see that the $\varphi$ function plays a similar role to a cosmological constant, which even setting $\Lambda_0=0$ could explain dark energy, if one fixes it to be equal to the observed value $\Lambda_c=10^{-52} \,\text{m}^{-2}$. 
For example, we can make the following ansatz
\begin{equation}
    \varphi\,=\,\mathrm{g} \left(\frac{m}{\Lambda}\right)^{2n} \lambda^{-2}_C\,,
\end{equation}
where $\mathrm{g}$ is some dimensionless coupling constant, and $\lambda_C$ is the Compton wavelength associated to the mass scale $m$. 

As a speculative exercise we can take as a reference mass scale the proton one, $1$ GeV, and notice that $ \lambda_C\approx 10^{-15}$ m, so that, if one imposes $\varphi=\Lambda_C$, then 
\begin{equation}
    \Lambda\,\approx\, ( 10^{82}\mathrm{g})^{1/2n} \,\text{GeV}\,.
\end{equation}
In this case, taking $n=2$ and $\mathrm{g}\approx10^{-6}$, one gets for the Lorentz deformation scale $\Lambda\approx 10^{19}$ GeV, i.e., the Planck scale. 

While there is an obvious high degree of arbitrariness in such a choice of scales, we just want to stress that a possible consequence of the cotangent bundle geometries taken here into account is the induction of a mass dependent cosmological constant, where the mass scale $m$ could be an averaged scale over the standard model mass scales or associated to new physics (e.g. super-symmetry scale). Finally, note that the function $\varphi$ is the only way in which momentum dependence of the Einstein equations can arise.

\section{Conclusion}
\label{sec:conclusions}
In this paper, we studied the Rindler spacetime with a momentum dependent metric that takes into account a relativistic deformed kinematics in curved spacetimes. While previous work provided a description of such spacetime from a quantum theory viewpoint~\cite{Arzano:2017opw}, here we focused on its geometrical realization. We started by deriving the trajectories of uniformly accelerated observers in a momentum deformed metric, confirming they are hyperbolae corresponding to a boost Killing flow, though now depending on the mass of the particle probing the spacetime. This allowed us to consider the horizon of the Rindler wedge, hence the very Rindler causal structure, undeformed with respect to the GR case. 

Albeit a QFT based on the symmetries of DSR is still missing, we assumed that the Bisognano--Wichmann still holds in this framework, so as to define the temperature of the Rindler wedge from the Unruh temperature, modified by the momentum dependence of the metric, with a factor solely depending on the mass of the observer, as it was the case of the deformed uniformly accelerated trajectories. Such a temperature is the same one of GR for the privileged momentum metric obtained in~\cite{Relancio:2020rys,Relancio:2021asx}.

Starting from the thermal Rindler wedge description, we considered the derivation of the Einstein equations of GR as a thermodynamic equation of state, as first proposed in \cite{Jacobson:1995ab}, in our momentum deformed setting. By reproducing the original derivation, we obtained deformed Einstein's equations with a momentum dependent metric. Remarkably, these equations turned out to be same one proposed in~\cite{miron2001geometry}, but with a different definition of the Riemann tensor and a new term depending on the mass of the particle probing spacetime. The latter equations are in fact (up to the new term) the same ones obtained in previous work by some of the present authors~\cite{Relancio:2020rys}, when considering the proposal of~\cite{miron2001geometry} together with imposing the conservation of the stress-energy tensor. We showed that such a new term can be regarded as a cosmological constant, as discussed in the previous section with some numerical estimations. 

The thermodynamic derivation of the Einstein equations happens to support the idea that the energy-momentum tensor is conserved only in the privileged momentum metric previously obtained by some of the present authors in~\cite{Relancio:2020rys,Relancio:2021asx}. As discussed in those papers, there is a long debate in the DSR community about the possible fact that different basis could represent different physics. In~\cite{Relancio:2020rys}, a degeneracy on the choice of momentum variables for flat spacetime was shown to exist, which is broken when introducing a curvature on spacetime. In this paper, we observe also this breaking of the degeneracy when considering curved spacetimes, pointing to the possibility that a privileged momentum basis, the one given by~\eqref{eq:conformal_metric}, may in fact exist.

\section*{Acknowledgments}
SL acknowledge funding from the Italian Ministry of Education and Scientific Research (MIUR) under the grant PRIN MIUR 2017-MB8AEZ.   JJR acknowledges support from the Unión Europea-NextGenerationEU (``Ayudas Margarita Salas para la formación de jóvenes doctores''). This work has been partially supported by Agencia Estatal de Investigaci\'on (Spain)  under grant  PID2019-106802GB-I00/AEI/10.13039/501100011033.
 The authors would also like to thank support from the COST Action CA18108.

\end{document}